\documentclass[aps,prl,twocolumn,groupedaddress,showpacs,floatfix,superscriptaddress,longbibliography]{revtex4-2}
\usepackage[plainpages=false,pdfpagelabels,colorlinks=true,linkcolor=red,urlcolor=blue,citecolor=blue,pdftitle={Title},pdfauthor={},pdfdisplaydoctitle=true,pdfduplex=DuplexFlipLongEdge]{hyperref}
\usepackage{siunitx}
\usepackage{mhchem}
\usepackage{epsfig}
\usepackage{graphicx}
\usepackage[utf8]{inputenc}
\usepackage{amsmath,amssymb}
\usepackage[dvipsnames]{xcolor}

\usepackage[normalem]{ulem}
\usepackage{tabularx}
\usepackage{xcolor}

\begin{document}

%% \title{Effect of Dilution on the Half-Filled Holstein Model\\
\title{Lattice-Distortion-Mediated Proton Pairing and Trapping in Solid State Oxides}
\author{Hang Ma}
\affiliation{School of Physics and Astronomy, Beijing Normal University, Beijing 100875, China\\}
\author{Jiajun Linghu}
\email{linghujiajun@chd.edu.cn}
\affiliation{ Department of Applied Physics, Chang'an University, Shaanxi, Xi'an, China\\}
\author{Nannan Han}
\affiliation{State Key Laboratory of Flexible Electronics (LOFE) \& Institute of Flexible Electronics (IFE) , Northwestern Polytechnical University, 127 West Youyi Road, Xi'an, 710072, China\\}
\author{Ying Liang}
\email{liangying@hebtu.edu.cn}
\affiliation{College of Physics, Hebei Normal University, Shijiazhuang 050024, China}
\affiliation{School of Physics and Astronomy, Beijing Normal University, Beijing 100875, China\\}
\affiliation{Key Laboratory of Multiscale Spin Physics (Ministry of Education), Beijing Normal University, Beijing 100875, China\\}
\author{Yiyang Sun}
\affiliation{State Key Laboratory of High Performance Ceramics, Shanghai Institute of Ceramics, Chinese Academy of Sciences, Shanghai 201899, China\\}
\author{Tianxing Ma}
\email{txma@bnu.edu.cn}
\affiliation{School of Physics and Astronomy, Beijing Normal University, Beijing 100875, China\\}
\affiliation{Key Laboratory of Multiscale Spin Physics (Ministry of Education), Beijing Normal University, Beijing 100875, China\\}
\author{Zhi-Peng Li}
\email{iamzpli@nwpu.edu.cn}
\affiliation{State Key Laboratory of Flexible Electronics (LOFE) \& Institute of Flexible Electronics (IFE) , Northwestern Polytechnical University, 127 West Youyi Road, Xi'an, 710072, China\\}

\begin{abstract}
Experiments have evidenced proton pairing in Y-doped BaZrO$_3$. However, the nature of proton pairing and its impact on conduction remain insufficiently understood theoretically. Here, through quantitative computational analysis of proton-proton interactions in Y-doped BaZrO$_3$, we identify lattice-distortion-mediated elastic interaction as the key factor determining whether two protons form a stable pair or exhibit net repulsion. When a proton resides at an inward-bending distortion site induced by another proton, the resulting net repulsive interaction leads to an unstable configuration. In contrast, the proton tends to be trapped at a nearby outward-bending site that favors the formation of a stable proton pair. Moreover, the site where the two protons form the lowest-energy configuration also corresponds to a proton trapping site. By calculating the long-range diffusion pathways accessible to protons under different local environments in both single- and two-proton cases, we find that the range of rate-limiting barriers is 0.24-0.45 eV for two-proton conduction and 0.19-0.39 eV for single-proton conduction. The higher and more experimentally consistent barriers in the two-proton pathways indicate that the proton trapping effect induced by pairing hinders proton conduction. Our study elucidates the multi-proton diffusion mechanism, providing a theoretical foundation for the experimental design of electrolytes with enhanced proton conductivity.  
\end{abstract}

\date{Version 16.0 -- \today}

\maketitle

%%%%%%%%%%%%%%%%%%%%%%%%%%%%%%%%%%%%%%%%%%%%%%%%%%%%%%%%%%%%%%%%%%%%%%%%
%%%%%%%%%%%%%%%%%%%%%%%%%%%%%%%%%%%%%%%%%%%%%%%%%%%%%%%%%%%%%%%%%%%%%%%%

\section{Introduction}
Amid the growing global demand for clean energy, solid oxide fuel cells (SOFCs) have attracted significant attention due to their high electrical efficiency ($>$50\%) and overall energy utilization exceeding 80\%\cite{XU2022115175,BICER20203670}. Compared to conventional oxygen-ion-conducting SOFCs that require high operating temperatures above 800~\si{\celsius}\cite{TIMURKUTLUK20161101,doi:10.1021/acsami.3c09025}, proton-conducting SOFCs have emerged as promising alternatives because of their high proton conductivity at interediate temperatures (400-600~\si{\celsius})\cite{ding2020self}. This enhanced conductivity is primarily attributed to the low activation energy for proton diffusion in the electrolyte. Since the proton diffusion rate is a key factor determining the conductivity, understanding proton diffusion mechanisms is crucial for the design and discovery of high-performance proton-conducting electrolytes.

Most proton-conducting electrolytes are transition metal oxides with the ABO$_3$ perovskite structure, such as II-IV perovskites (BaCeO$_3$, BaZrO$_3$, SrCeO$_3$, etc.)\cite{kannan2013chemically,B902343G,FABBRI20101043,https://doi.org/10.1111/jace.15946,Münch01041999} and III-III perovskites (YbCoO$_3$, SmNiO$_3$, SrRuO$_3$, etc.)\cite{LINGHU20252922,zhou2016strongly,li2020reversible}. In conventional II-IV ABO$_3$ systems, proton conduction is enabled by doping trivalent cations at the tetravalent B-site which promotes the formation of oxygen vacancies\cite{zhang2018insight,liu2020microscopic}. Under humid conditions, these vacancies incorporate hydroxyl groups from dissociated water, while the accompanying proton binds to a nearby oxygen site\cite{C8CP07632D}. The resulting protons bound to oxygen ions diffuse via the Grotthuss mechanism, which involves rotation around a single oxygen site and transfer between adjacent sites\cite{AGMON1995456,KREUER2000149}. A widely studied proton-conducting system is Y-doped BaZrO$_3$. Doping with trivalent cations such as Y$^{3+}$ enhances the proton conductivity of the BaZrO$_3$ system\cite{annurev:/content/journals/10.1146/annurev.matsci.33.022802.091825}. However, excessive Y doping, while increasing the proton concentration, can also introduce trapping regions that hinder proton diffusion\cite{FABBRI20101043}. Experimental and molecular dynamics (MD) simulation studies have reported activation energies of 0.42~eV\cite{raiteri2011reactive}, 0.43~eV\cite{annurev:/content/journals/10.1146/annurev.matsci.33.022802.091825}, 0.44~eV\cite{friman2013molecular}, 0.45~eV\cite{doi:10.1021/jp801082q}, and 0.47~eV\cite{KREUER1999285} for Y doping concentrations of 0.46\%, 10\%, 11\%, 12.5\%, and 20\%, respectively. This trend is commonly attributed to the increasing trapping effects with higher dopant concentrations, which hinder proton diffusion and thus raise the activation energy\cite{FABBRI20101043}. However, the increase in Y doping is also accompanied by an increase in proton concentration, and proton-proton correlations may also influence the resulting activation energy. 

Both proton rotation and transfer are thermally activated processes\cite{braun2017experimental}. At elevated temperatures, lattice vibrations involving outward bending of the O-B-O bond facilitate the breaking of hydrogen bonds between the proton and acceptor oxygen, enabling rotation easily around the A-site ion (The O-B-O angle denotes the bond angle between the oxygen covalently bonded to the proton, the B-site cation, and the adjacent oxygen that forms a hydrogen bond with the proton.). In contrast, inward bending modes promote the formation of nearly linear and strong hydrogen bonds\cite{10.1063/1.3122984,KREUER1995157,MUNCH2000183,JING2020227327,ma2025hydrogenbondstrengthdictates,rfs5-bnwt}, which assist the proton transfer to the acceptor oxygen barrierlessly\cite{PhysRevLett.102.075506}. Thus, long-range proton diffusion occurs through coupling with lattice dynamics\cite{KREUER2000149,MUNCH2000183,PhysRevLett.102.075506,SAMGIN2000291,du2020cooperative}. In line with this mechanism, quasielastic neutron scattering experiments on phonon vibration modes during proton transfer in the proton-conducting perovskite system Y-doped BaZrO$_3$ suggest a protonic polaron mass twice that of a free proton, providing evidence for proton pairing\cite{du2020cooperative}. This result contrasts with the expected Coulombic repulsion between protons, thereby calling for a comprehensive investigation of the underlying pairing mechanism. In superionic conductors, cooperative diffusion involving multiple lithium ions has been extensively studied and experimentally confirmed\cite{he2017origin,10.1063/1.4737397,doi:10.1021/cm303542x}. Whether a similar proton-pair diffusion mechanism exists in proton-conducting electrolytes, and whether multi-proton motion could facilitate charge transport, remain open questions.

In this study, through quantitative analysis of the interactions between two protons in Y-doped BaZrO$_3$ based on density functional theory calculation (DFT), we reveal that lattice-distortion-mediated elastic interaction governs whether two protons form a stable pair or exhibit net repulsion. Due to proton pairing, protons tend to localize at two types of trapping sites: (i) sites adjacent to inward-bending, net repulsive positions induced by excess protons, and (ii) sites corresponding to the lowest-energy proton pair configuration. The calculation the long-range diffusion pathways in both single- and two-proton cases reveal that the limiting barrier is higher and closer to the experimental activation energy in the two-proton pathways, indicating that the proton trapping effect induced by pairing hinders proton conduction. Our study clarifies multi-proton diffusion and supports the design of high-conductivity electrolytes.
%%%%%%%%%%%%%%%%%%%%%%%%%%%%%%%%%%%%%%%%%%%%%%%%%%%%%%%%%%%%%%%%%%%%%%%%
%%%%%%%%%%%%%%%%%%%%%%%%%%%%%%%%%%%%%%%%%%%%%%%%%%%%%%%%%%%%%%%%%%%%%%%%

\section{Method}
%%{\color{blue}
All density functional theory calculations were performed by the Vienna Ab Initio Simulation Package (VASP) with projector augmented wave (PAW) method\cite{KRESSE199615,PhysRevB.59.1758,PhysRevB.54.11169}. The Perdew-Burke-Ernzerhof (PBE) based generalized gradient approximation (GGA)\cite{PhysRevLett.77.3865} exchange-correlation functional is used. The cutoff energy of the plane wave basis was set to 520~eV. All atoms are relaxed using the conjugate gradient algorithm until the forces on each atoms are less than 0.01~eV/\AA. The Monkhorst-Pack method\cite{PhysRevB.13.5188} was employed for k-point sampling, with a spacing of 0.028~\AA$^{-1}$. Proton migration barriers were calculated in a $2\times2\times4$ supercell containing 80 atoms using the climbing image nudged elastic band (CI-NEB) method\cite{10.1063/1.1329672}. Structural optimization of initial and final states for both proton transfer and rotation were performed with relaxed lattice constants, while optimization for each intermediate image was restricted to atomic coordinate relaxations with fixed lattice parameters. The residual force were converged to less than 0.03~eV/\AA. All calculations were performed with spin polarization considered. After relaxation, the magnetic moment of hydrogen becomes zero due to the covalent bonding between H and O, and the entire system is nonmagnetic.

\section{Results and Discussion}
\subsection*{A. The configuration of proton pairs}

\begin{figure}[htbp]
\centering
\includegraphics[scale=0.27]{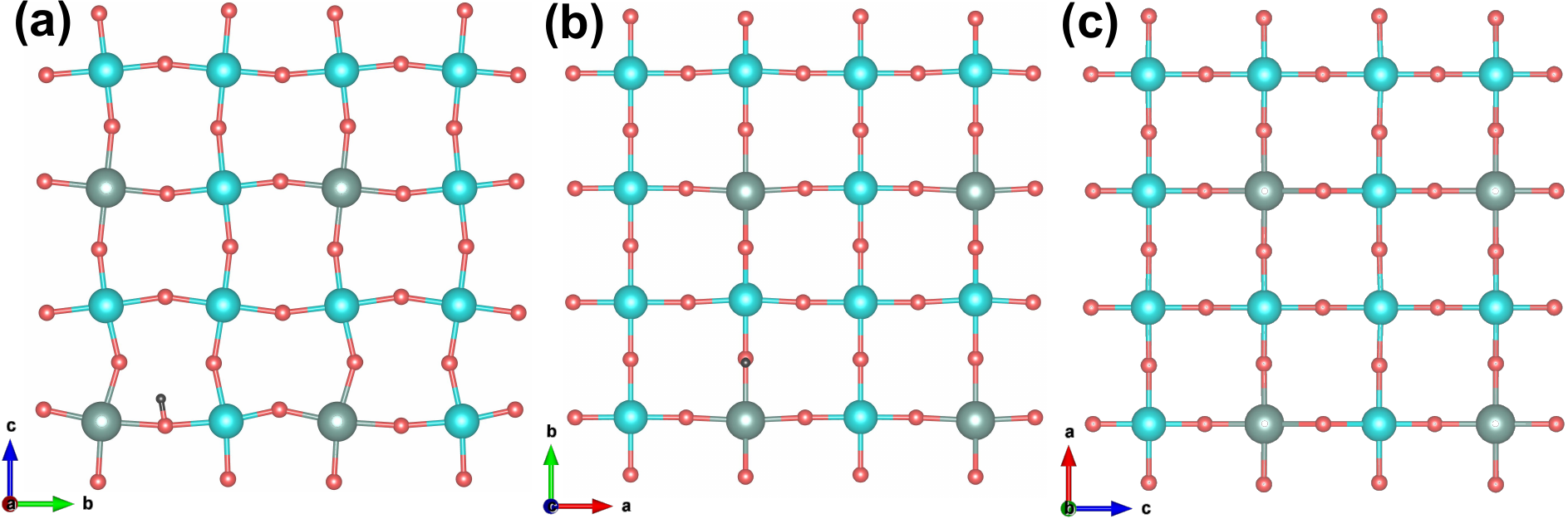}
\caption{Lattice distortions induced by a proton in the (a) bc, (b) ba, and (c) ac planes. Cyan-blue, gray-green, and pink spheres represent Zr, Y, and O atoms, respectively, while the black sphere denotes the first fixed proton.}
\label{Fig1}
\end{figure}

\begin{figure}[htbp]
\centering
\includegraphics[scale=0.38]{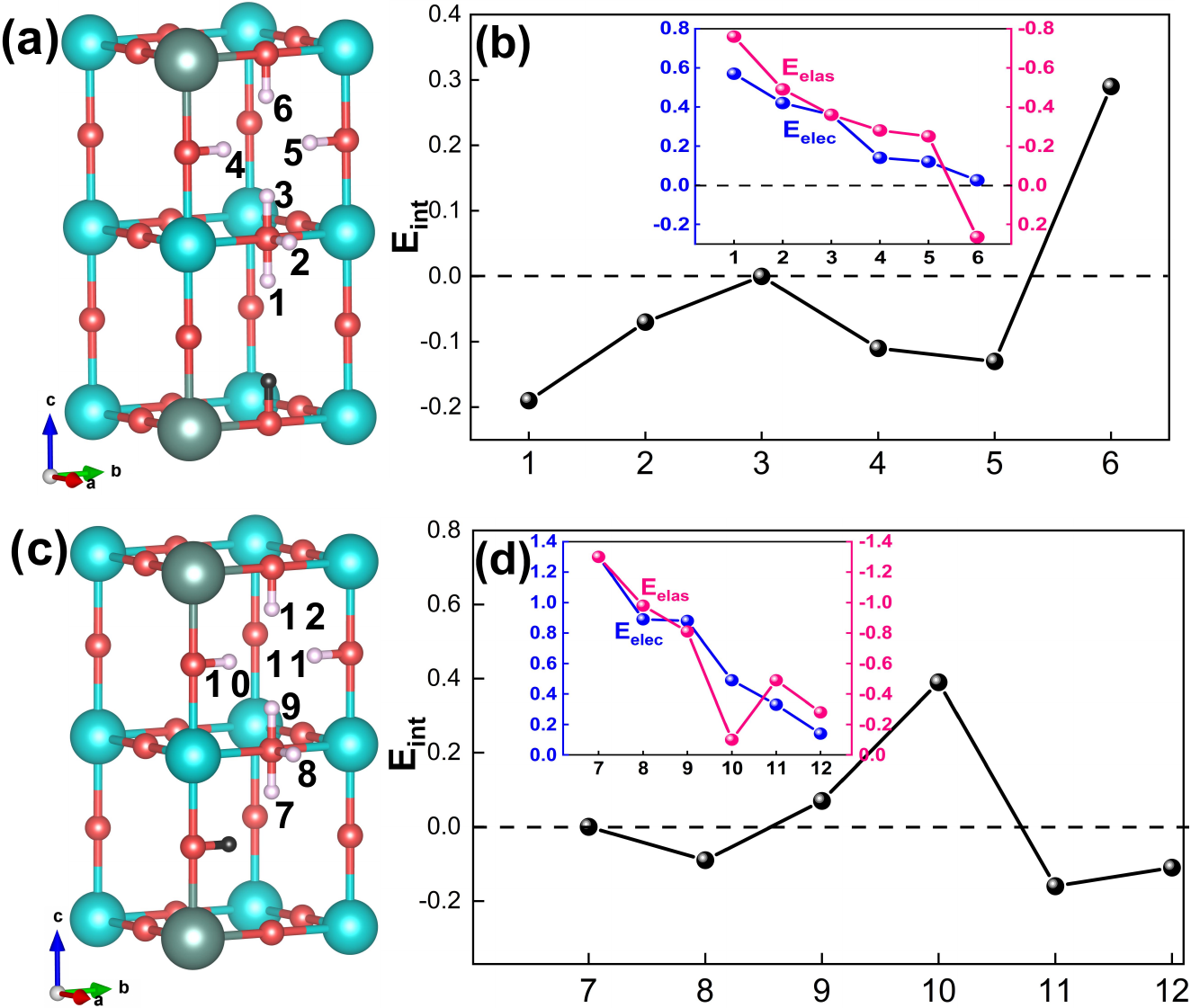}
\caption{(a,c) Schematic illustration of two-proton configurations with proton 1 fixed at the Y-adjacent oxygen site along the b-axis (a) and c-axis (c). The black sphere represents the fixed proton 1, and the white sphere represents proton 2 at different positions. (b,d) Proton-proton interaction energy $E_{int}$, and electrostatic repulsion energy $E_{elec}$ (blue line of insert) and elastic interaction energy $E_{elas}$ (pink line of insert) for the configurations in (a) and (c), respectively. }
\label{Fig2}
\end{figure}

Based on the experimental and theoretical results reported by Maria A. Gomez et al\cite{du2020cooperative,article}, a proton can induce local lattice distortions that stabilize a second proton at a nearby site through the formation of a strong hydrogen bond, thereby overcoming their mutual repulsion and forming a proton pair. To identify which site configurations lead to a net attractive interaction between two protons, we conducted a more comprehensive investigation beyond the previously studied case where both protons are localized within the same Y-induced trapping region\cite{du2020cooperative,article,inorganics11040160}. A single proton induces antiferrodistortive octahedral rotation along the b and c directions within its own plane (Fig.~\ref{Fig1}). Such antiphase rotational distortions shorten the distances between the proton and surrounding oxygen ions, thereby stabilizing the proton\cite{PhysRevB.107.134102}, as confirmed by $\Gamma$-point phonon calculations showing no imaginary frequencies. This also indicates that one proton can influence another proton located in the same plane through lattice distortion\cite{article}, whereas in the non-coplanar configurations (13 and 14) discussed in Section 1 of the Appendix, the lattice-distortion interaction is much weaker. Therefore, we mainly focus on the two-proton configurations where proton 2 is located at the coplanar oxygen site with proton 1. We first investigated a $2\times2\times4$ BaZrO$_3$ supercell with 12.5\% uniformly distributed Y doping, following the concentration used in the work of Maria A. Gomez et al.\cite{du2020cooperative}. Proton 1 (black sphere in Fig.~\ref{Fig2}(a,d)) was fixed at one of the two nearest oxygen sites adjacent to a Y dopant. Proton 2 (white sphere) was then placed at various oxygen sites corresponding to configurations 1 through 12, where it either shares the same Y-induced trapping region with proton 1 or resides in an adjacent trapping region(The trapping region extends up to the next-nearest-neighbor sites of Y\cite{bevillon2014dopant,PhysRevB.76.054307}).

We first considered the case where proton 1 is fixed at the Y-adjacent oxygen site along the b-axis direction, while proton 2 occupies one of the sites labeled 1 through 6 (Fig.~\ref{Fig2}(a)). To quantitatively evaluate whether there exists a net attractive or repulsive interaction between two protons, we calculated their interaction energy. The total energy of a given two-proton configuration is denoted as $E_{tot}$. As a reference energy $E_{ref}$, we used the total energy of two independent protons placed far apart but within the same Y-induced trapping environment, such that proton-proton interactions are negligible. The interaction energy is then defined as $E_{int} = E_{tot} - E_{ref}$ (The detailed procedure and data is in Fig.~\ref{FigA1} and Table.~\ref{TabA1}). As shown in Fig.~\ref{Fig2}(b), the interaction energy $E_{int}$ is negative for most positions of proton 2, indicating the formation of proton pairs with net attractive interaction. In contrast, when proton 2 is located at site 6, $E_{int}$ is positive, suggesting a net repulsive interaction with proton 1. We decompose the proton-proton interaction energy $E_{int}$ into two components: the electrostatic repulsion $E_{elec}$ between the two protons, and the elastic interaction $E_{elas}$ arising from lattice distortions induced by the excess protons (According to the Bader analysis, the two protons exhibit charges of +0.57 and +0.6, resulting in a repulsive electrostatic interaction.). A negative $E_{int}$, corresponding to a proton pair, indicates that the favorable, attractive elastic interaction overcomes the Coulomb repulsion. But the absence of a proton pair when proton 2 occupies site 6 requires further quantitative investigation. The electrostatic repulsion energy, $E_{elec}$, is calculated as the difference between the total energy $E_{tot}$ of the current two-proton configuration and the energy $E_{ref}$ of the corresponding independent proton reference system without relaxaion. The elastic interaction energy $E_{elas}$ is then obtained by subtracting $E_{elec}$ from the total interaction energy $E_{int}$\cite{PhysRevB.76.054307,C3TA12870A}. As shown in the insert of Fig.~\ref{Fig2}(b), when proton 2 occupies site 6, the positive value of $E_{elas}$ indicates a repulsive elastic interaction with proton 1, implying that the lattice distortion induced by proton 1 destabilizes proton 2 at site 6.  Such an elastic interaction, unable to compensate the Coulombic repulsion, is the main contributor to the relatively large $E_{int}$. Similarly, we also calculated the proton-proton interaction when proton 1 is fixed at the Y-adjacent oxygen site along the c-axis, and proton 2 occupies sites 7 through 12, as shown in Fig.~\ref{Fig2}(c). Among these configurations, only when proton 2 is located at site 10 does the interaction energy $E_{int}$ become significantly positive and larger compared to other sites (Fig. \ref{Fig2}(d)), indicating a net repulsive interaction. The consistency between the trends of the elastic interaction $E_{elas}$ and the total interaction $E_{int}$—with their non-monotonic variation as the proton-proton distance increases (see Table 1)—indicates that the inability to form a proton pair at this site is again attributed to the near-zero $E_{elas}$ value(the insert of Fig.~\ref{Fig2}(d)). Therefore, the elastic interaction is the key factor determining whether a two-proton configuration forms a stable pair or exhibits mutually net repulsion.

\begin{figure}[htbp]
\centering
\includegraphics[scale=0.3]{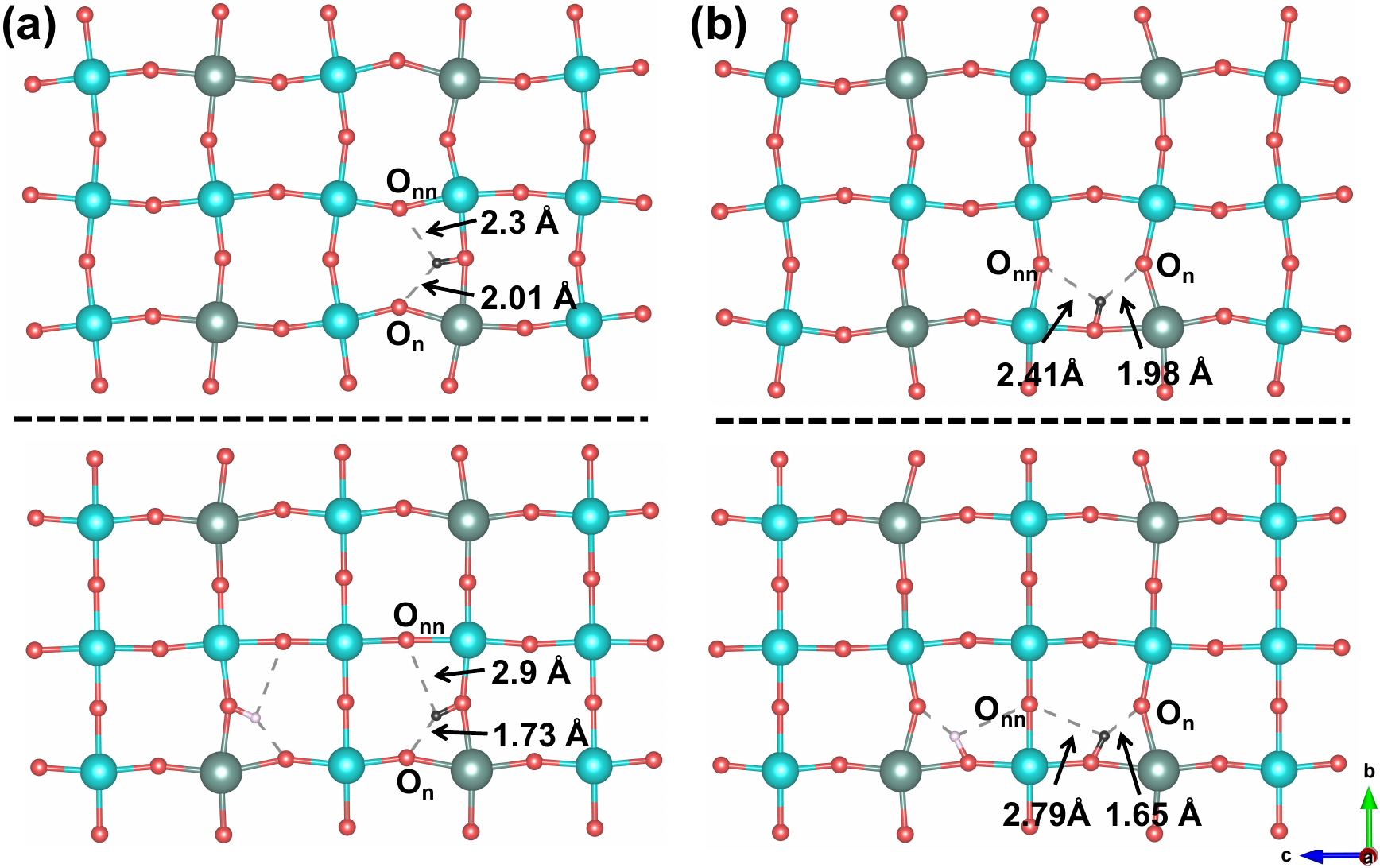}
\caption{(a) Lattice distortion in the plane induced by a proton at the nearest-neighbor site of Y along the b-axis (top), and the modified distortion when a second proton is added at the repulsive site 6 (bottom). (b) Similar distortions for a proton at the Y-adjacent site along the c-axis (top), and with an additional proton at repulsive site 10 (bottom). }
\label{Fig3}
\end{figure}

\begin{table*}[htbp]
\centering
\caption{The $H-O_{nn}$, $H-O_{nn}$ bond lengths and the distance between the two protons $H-H$ for all two-proton configurations shown in Fig.~\ref{Fig2}(a,d), where O$_n$ and O$_{nn}$ denote the nearest and next-nearest oxygen neighbors of the proton, respectively.}
\label{Tab1}
\renewcommand{\arraystretch}{1.5}
\setlength{\tabcolsep}{7pt}
\begin{tabular}{|c|c c c c c c c c c c c c|}
\hline
site & 1 & 2 & 3 & 4 & 5 & 6 & 7 & 8 & 9 & 10 & 11 & 12 \\ \hline
$H-O_{nn}$~(\AA) & 2.19 & 2.29 & 2.28 & 2.28 & 2.18 & 2.9 & 2.06 & 2.29 & 2.4 & 2.79 & 2.4 & 2.28\\
$H-O_{n}$~(\AA) & 1.99 & 1.98 & 1.96 & 1.99 & 2.1 & 1.73 & 1.73 & 2.1 & 1.84 & 1.65 & 1.87 & 1.98 \\$H-H$~(\AA) & 2.61 & 3.49 & 4.69 & 5.65 & 5.9 & 6.21 & 2.15 & 2.57 & 3.36 & 5.13 & 5.35 & 5.63 \\
\hline
\end{tabular} 
\end{table*}

The elastic interaction between two protons essentially arises from the lattice distortion induced by one proton, which modifies the electrostatic interactions between the other proton and the surrounding lattice ions. Therefore, we analyze the bond lengths between the proton and nearby oxygen ions to explain why repulsive elastic interactions occur when proton 2 is located at sites 6 and 10. Furthermore, to evaluate the correlation between these bond lengths and the lattice-distortion interaction, we measured and computed $\frac{1}{H-O_n}+\frac{1}{H-O_{nn}}$, representing the dominant electrostatic attraction between the proton and its nearest oxygen O$_n$ and next-nearest oxygen O$_{nn}$, along with the corresponding two-proton elastic interaction energy $E_{elas}$ for all dopant and proton configurations considered in this study. Using the Spearman correlation coefficient\cite{hauke2011comparison}, which characterizes the monotonic relationship between two variables, we found a general negative Spearman correlation, with the strongest correlation coefficient reaching -0.829 and a p-value of 0.04, indicating statistical significance. The low values of $\frac{1}{H-O_n}+\frac{1}{H-O_{nn}}$ at sites 6 and 10 are primarily due to the elongated $H-O_{nn}$ distances. As shown in Fig.~\ref{Fig3}(a,b) and Table.~\ref{Tab1}, the $H-O_{nn}$ bond length increases from 2.3~\AA\ in the single-proton case to 2.9~\AA\ at replusive site 6. For site 10, the $H-O_{nn}$ distance reaches 2.79~\AA, compared to 2.41~\AA\ in the corresponding single-proton configuration. Meanwhile, the $H-O_{nn}$ bond lengths in other configurations that form stable proton pairs are all shorter than those in the corresponding single-proton systems. The reason is that when a second proton is placed at the repulsive site 6 or 10, the phase of its induced octahedral rotation opposes that of the first proton. This phase mismatch disrupts the original stabilizing antiferro rotation pattern, elongates the $H-O_{nn}$ bond, and leads to a repulsive elastic interaction. Furthermore, phonon calculations for these two configurations yield imaginary frequencies of -82.81~$cm^{-1}$ and -66.36~$cm^{-1}$, respectively. The absolute values of these frequencies correspond to octahedral rotational modes\cite{ZEUDMISAHRAOUI2013195}, indicating that the proton resides in a locally unstable, high-energy state. 

A common feature of these two repulsive configurations is that proton 2 occupies an oxygen site with inward-bending distortion induced by proton 1 along the direction of proton 2, resulting in the mismatch of rotation phase. According to the pattern of antiferro-type octahedral rotations, the neighboring site of this inward-bending position is outward-bending. When proton 2 resides at this outward-bending site, the two protons induce octahedral rotations with the same phase, resulting in a favorable elastic interaction that overcomes the Coulomb repulsion and forms a stable low-energy proton pair. As a result, proton 2 tends to be trapped at the site adjacent to the repulsive one. It is worth emphasizing that our analysis is based on the $2\times2\times4$ supercell. To eliminate artifacts arising from periodic boundary conditions, we also performed calculations using $2\times2\times6$ and $4\times4\times4$ supercells for validation (Fig.~\ref{FigA2}). In all cases, the proton located at the inward-bending repulsive site was found to disrupt the antiferrodistortive octahedral rotation induced by the other proton.

Since the ionic radius of Y is larger than that of Zr, high-density Y distributions are more likely to induce octahedral distortions, resulting in inward or outward-bending sites. To rule out the influence of uniformly distributed Y dopants on the net repulsive two-proton configurations, we also examined configurations under a non-uniform 12.5\% Y doping condition (Fig.~\ref{FigA3}). As in the case of uniformly distributed Y, net repulsive interactions occur between two protons when proton 2 occupies sites 6$'$ and 10$'$. The origin of this repulsion is that, at these sites, proton 2 sits on inward-bending positions induced by proton 1, resulting in octahedral rotations with opposite phases. This mismatch partially cancels the lattice distortions each other, elongating the proton-oxygen distances and resulting in repulsive lattice-distortion interactions. Consequently, $E_{elas}$ cannot compensate for $E_{elec}$, leading to a positive $E_{int}$.

\subsection*{B. The conduction of proton pairs}

\begin{figure*}[htbp]
\centering
\includegraphics[scale=0.43]{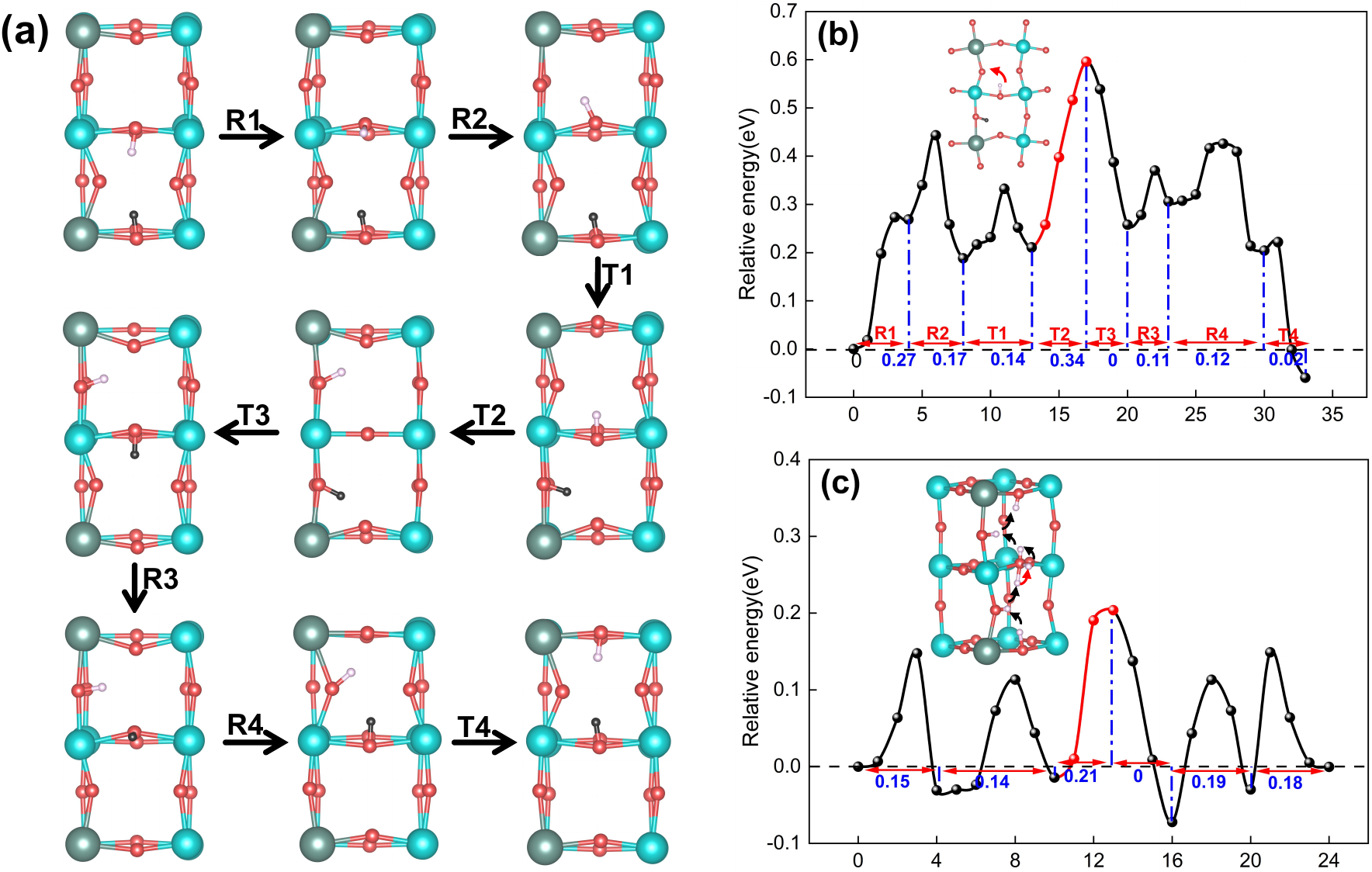}
\caption{(a) Schematic of the minimum energy pathway for two-proton migration under uniform Y doping. (b) Energy barrier along the pathway shown in (a); the inset illustrates the rate-limiting step(red line). (c) Single-proton diffusion pathway (inset) and energy barrier corresponding to (a), with the red line indicating the rate-limiting step.}
\label{Fig4}
\end{figure*}

Among all the two-proton configurations shown in Fig.~\ref{Fig2}, the strongest net attractive interaction occurs when proton 2 is located at site 1. This configuration also exhibits the lowest total energy, indicating that site 1 serves as another trapping site in addition to those adjacent to the repulsive positions. This result is consistent with previous studies, which identified the lowest-energy two-proton configuration as one in which both protons are trapped within the same plane of a Y-centered unit cell and form hydrogen bonds with the same oxygen ion\cite{du2020cooperative,article,inorganics11040160}. To further examine whether two protons tend to be trapped at the sites corresponding to the lowest-energy proton pair configuration identified by our DFT calculations, we performed 10 ps ab initio molecular dynamics (AIMD) simulations for $3 \times 3 \times 3$ BaZrO$_3$ system with 3.7\% Y doping containing two protons at 700~K and 1300~K. The static radial distribution function between the two protons was computed based on the van Hove correlation function at zero time lag (computational details are given in the Appendix). The doping concentration was chosen to probe the practically relevant trapping effect of a single Y dopant on two protons, while the two temperatures represent, respectively, the lower-to-intermediate operating range of proton-conducting fuel cells and a condition above the typical processing temperature, enabling us to assess the stability of the two-proton configuration near Y. As shown in Fig.~\ref{FigA4}, the peak positions of $G_d(r,0)$ at 700~K and 1300~K are located at approximately 2.3~\AA\ and 2.23~\AA, respectively, which are close to the proton-proton distance when proton 2 is at site 1. This indicates that two protons in the vicinity of a Y dopant tend to adopt the lowest-energy configuration, in which one proton is effectively trapped by both the Y dopant and the other proton. To verify whether protons undergo long-range diffusion in the form of closely spaced proton pairs, we performed CI-NEB calculations to determine the migration barrier along the c-axis, using the lowest-energy two-proton configuration as the initial state. Since proton conduction in Y-doped BaZrO$_3$ is understood to proceed via a series of rotations and transfers from one dopant trapping region to another\cite{10.1063/1.3122984,draber2020nanoscale,10.1063/1.3447377,10.1063/5.0039103}, the final state was chosen such that the two protons are located in the trapping region of a dopant adjacent to the dopant in the initial state. To better reflect realistic conditions, we calculated the long-range migration barriers of two-proton configurations under both uniform and non-uniform 12.5\% Y doping (Fig.~\ref{FigA5}).

\begin{figure*}[htbp]
\centering
\includegraphics[scale=0.43]{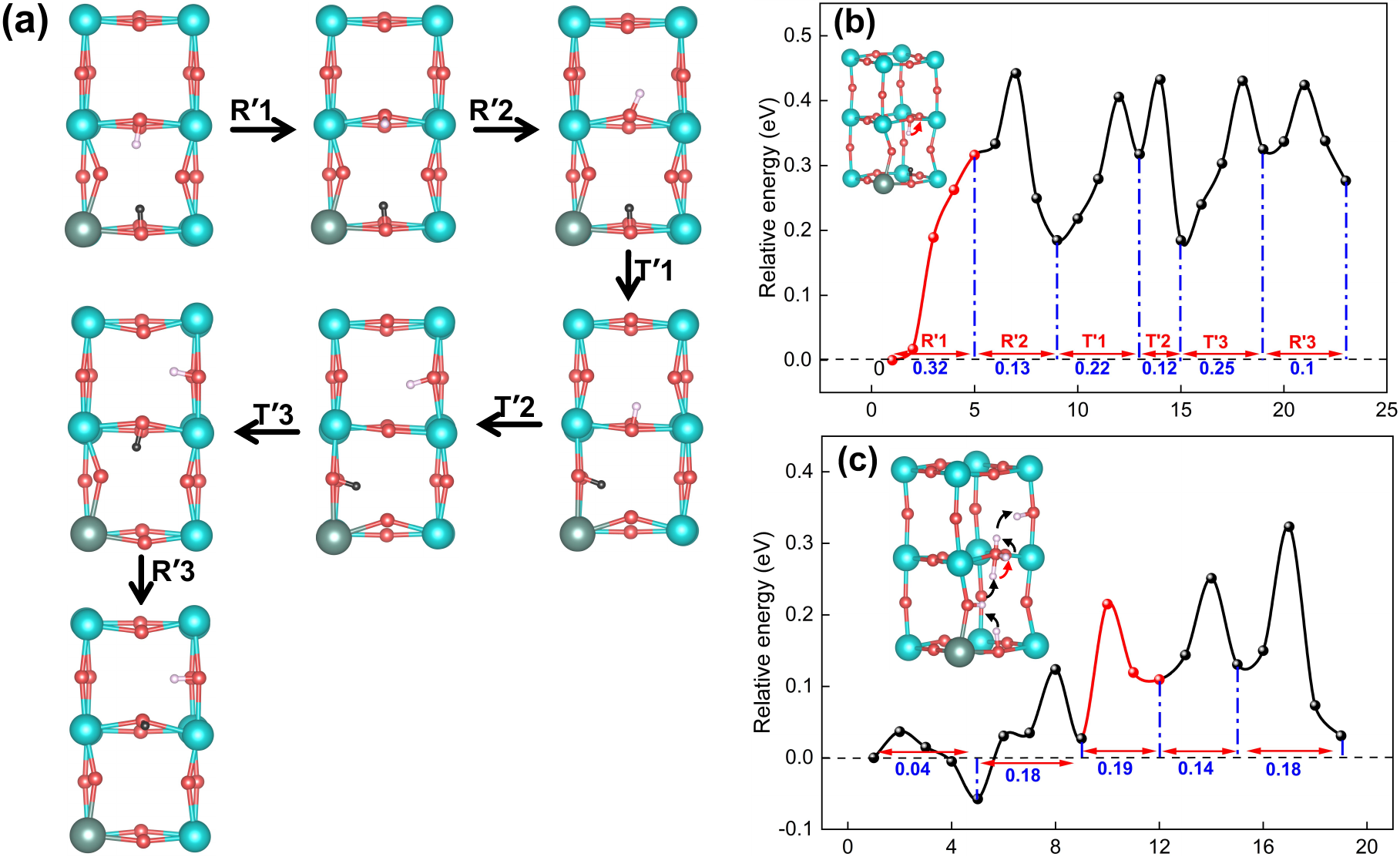}
\includegraphics[scale=0.4]{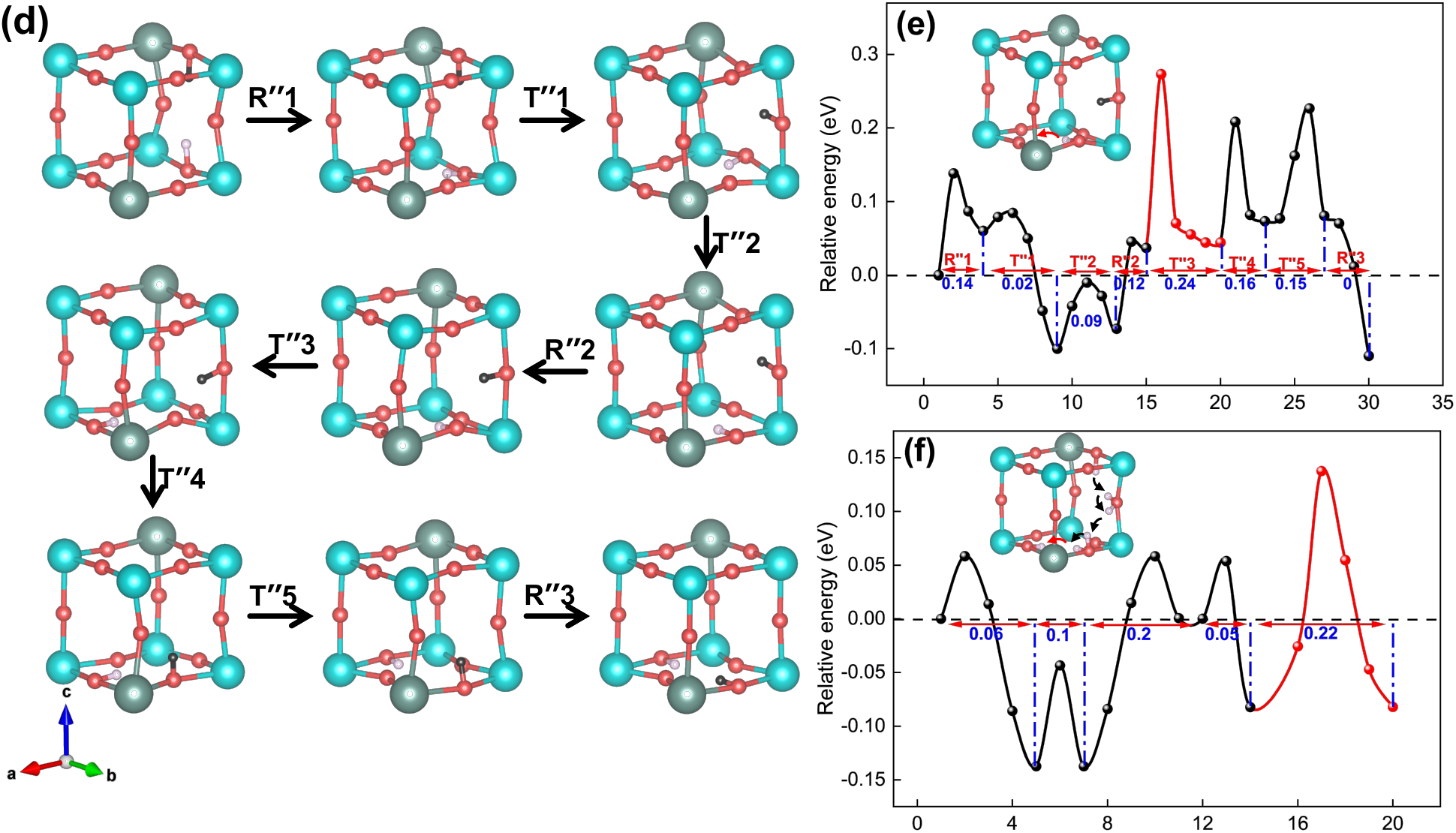}
\caption{(a) Schematic of the two-proton diffusion pathway in the low-density, non-uniform Y-doped region. (b) Energy barrier along the pathway shown in (a); the inset illustrates the rate-limiting step (red line). (c) Single-proton diffusion pathway (inset) and energy barrier corresponding to (a), with the red line indicating the rate-limiting step. (d-f) Two-proton and single-proton diffusion in the high-density, non-uniform Y-doped region.}
\label{Fig5}
\end{figure*}

Figure~\ref{Fig4}(a) shows the minimum energy pathway obtained from CI-NEB calculations under uniform Y doping (see Fig.~\ref{FigA6} and Table.~\ref{TabA3} for computational details and data). Proton 2 first escapes from the initial trapping region through two consecutive $90^\circ$ rotations (R1 and R2), with energy barriers of 0.27~eV and 0.17~eV, respectively. Then, proton 1 undergoes an intra-octahedral transfer (T1) around the Y dopant in the initial state, with a barrier of 0.14~eV. Subsequently, proton 2 and proton 1 perform intra-octahedral transfers (T2 and T3) around an intermediate Zr ion, bringing both protons closer to the final-state Y site, with barriers of 0.34~eV and 0~eV, respectively. Proton 1 then follows proton 2 to the trapping plane of the final state via two consecutive $90^\circ$ out-of-plane rotations (R3 and R4), with barriers of 0.11~eV and 0.12~eV. Finally, proton 2 completes the migration by undergoing an intra-octahedral transfer (T4) around the final-state Y dopant, with a barrier of 0.02~eV, bringing the system back to its lowest-energy configuration. The rate-limiting barrier along this pathway is 0.34~eV (red line in Fig.~\ref{Fig4}(b)), corresponding to T2, where proton 2 transfers into a net repulsive site(where the highest migration barrier along the pathway is considered the rate-limiting step). This step represents overcoming the trapping effect induced by the other proton, which accounts for the relatively high barrier. To examine the effect of proton pairing on conduction, we also calculated the single-proton migration barriers corresponding to the pathway in Fig.~\ref{Fig4}(a), as shown in Fig.~\ref{Fig4}(c). The rate-limiting barrier for the single-proton case is 0.21~eV, associated with the rotation of the proton out of the Y trapping plane (indicated by the red arrow in the inset). Since it is not subject to trapping by another proton, this barrier (0.21~eV) is lower than that of the R1 step in the two-proton case (0.27~eV). For the same reason, the highest two-proton barrier (0.34~eV) is also higher than the corresponding single-proton migration barrier (0.19~eV).

\begin{figure*}[htbp]
\centering
\includegraphics[scale=0.43]{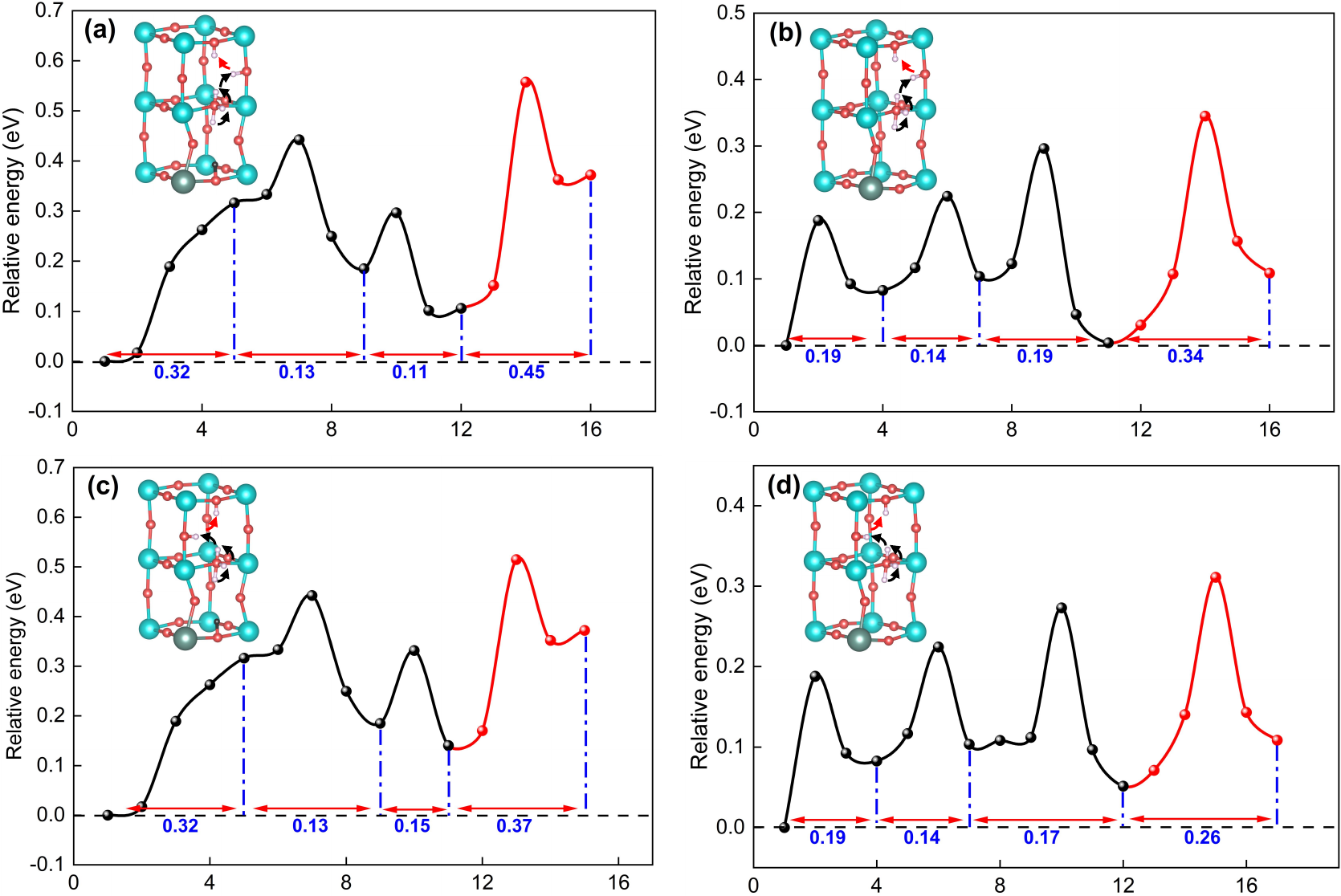}
\caption{(a,c) Energy barriers and schematic illustrations (insets) of two diffusion pathways for proton 2 moving away from the fixed proton 1. (b,d) Corresponding energy barriers and schematic illustrations (insets) for single-proton diffusion along the same pathways as in (a) and (c).}
\label{Fig6}
\end{figure*}

We categorized the non-uniform Y doping case into two environments-low-density (Fig.~\ref{Fig5}(a)) and high-density Y regions (Fig~\ref{Fig5}(d))—and performed calculations for each separately. We first examine the two-proton diffusion behavior in the low-density Y-doped region. Due to the large separation between neighboring dopants in this environment, we only calculated the pathway involving the escape of the two protons from the initial Y trapping region. As shown in Fig.~\ref{Fig5}(b), the rate-limiting barrier along this pathway is 0.32~eV, corresponding to step R$'$1, where proton 2 undergoes a $90^\circ$ rotation to escape from the initial low-energy state co-trapping by proton 1 and dopant. In the corresponding single-proton diffusion pathway shown in Fig.~\ref{Fig5}(c), the same rotational step constitutes the rate-limiting barrier. However, in the absence of trapping by another proton, the barrier is significantly lower (0.19~eV). In the high-density Y-doped region, the adjacent Y dopants are located at diagonal sites within the same unit cell. As a result, during the two-proton diffusion, the two protons occupy mutually perpendicular planes (the bc, ab, and ac planes containing the proton in Fig.~\ref{Fig5}(d)). The highest energy barrier along this pathway is 0.24~eV (Fig.~\ref{Fig5}(e)), corresponding to step T$''$3, which also represents the largest migration barrier step in the single-proton diffusion pathway under the same high-density doping condition. The corresponding single-proton barrier is 0.22~eV (Fig.~\ref{Fig5}(f)), and the barrier difference is negligible. Throughout the entire pathway, the two-proton migration barriers remain low and are nearly identical to those of the single-proton case. This observation further suggests that proton diffusion is primarily influenced by other protons within the same plane, with trapping effects manifesting under coplanar configurations. A high-density Y distribution across non-coplanar planes facilitates the spatial separation of protons, thereby reducing proton trapping effect that impedes diffusion and consequently enhancing proton conductivity.

In addition to the previously calculated diffusion pathway, where the two protons remain in close proximity, we also calculated pathways where proton 1 is fixed and proton 2 migrates away from proton 1 along positions shown in Fig.~\ref{Fig2}(a). Figures~\ref{Fig6}(a) and \ref{Fig6}(c) depict the energy barriers for proton 2 migrating via site 5 on the right and via site 4 on the left to reach site 6, respectively. The rate-limiting barriers along these two pathways correspond to proton transfers from sites with net attractive interactions (sites 5 and 4) to the net repulsive site 6, with barrier heights of 0.45~eV and 0.37~eV, respectively. Due to the proton trapping effect, these barriers are higher than the corresponding single-proton barriers-0.34~eV and 0.26~eV (Figs.~\ref{Fig6}(b) and \ref{Fig6}(d))-which also represent the rate-limiting barriers for single-proton diffusion along the pathways shown in Figs.~\ref{Fig6}(a) and \ref{Fig6}(c). For the interacting two-proton system, we have systematically explored nearly all possible migration scenarios and identified the highest energy barriers arising from proton trapping effects that hinder diffusion. For the single-proton case, to identify the highest energy barriers associated with Y-induced trapping as thoroughly as possible, we modified the CI-NEB pathway shown in Fig.~\ref{Fig2}(a)—which corresponds to diffusion along the Y trapping region—and instead computed a path where the proton migrates along the right side. In this case, the highest energy barrier is 0.39~eV, higher than that of the corresponding two-proton configuration. This barrier corresponds to the proton escaping from the Y trapping site, specifically through a transfer between two next-nearest-neighbor oxygen sites coordinated to the same Y dopant.

Based on the calculated diffusion pathways that incorporate trapping effects from both proton and low-valence dopants, the range of rate-limiting barriers for two-proton diffusion is 0.24-0.45~eV, while that for single-proton is 0.19-0.39~eV. The higher and more experimentally consistent two-proton barriers (compared to the reported value of 0.44~eV\cite{annurev:/content/journals/10.1146/annurev.matsci.33.022802.091825,Bohn2000}) correspond to steps in which one proton attempts to rotate or transfer at a site from the two types of trapping sites discussed in the first part of the paper. This clearly supports the existence of lattice distortion-mediated proton trapping effect that impede long-range proton diffusion. Due to the proton trapping effect, the highest rate-limiting barrier for two-proton diffusion is 0.45~eV, whereas the highest barrier for single-proton diffusion—arising from Y-induced trapping—is 0.39~eV. According to transition state theory\cite{RevModPhys.62.251,lin2024comparative}, the diffusion coefficient is proportional to the hopping rate, which depends exponentially on the activation energy. At an upper intermediate temperature of 600~\si{\celsius}, the 0.06~eV difference in activation energy leads to a diffusion coefficient ratio of approximately 0.45 between the two-proton and single-proton systems. Since the conductivity is proportional to the product of proton concentration and diffusion coefficient\cite{he2018statistical}, even after accounting for the doubled proton concentration in the two-proton system, the resulting conductivity ratio is 0.9, indicating that the conductivity remains lower in the two-proton case. Therefore, the increased activation barrier caused by proton trapping is sufficient to hinder proton conduction

\section{Summary}
In this work, to uncover the nature of proton pairing in Y-doped BaZrO$_3$, we perform a quantitative analysis of the interactions between two protons in various configurations. Our calculations reveal that the key factor determining whether two protons form a stable pair or exhibits net repulsion is the lattice-distortion-mediated elastic interaction between them. A single proton induces an antiferrodistortive octahedral rotation. When another proton occupies an inward-bending site caused by this distortion, the rotational phases induces by both protons cancel each other, giving rise to a lattice-distortion interaction that fails to compensate the Coulomb repulsion. As a result, a net repulsive interaction emerges, placing the system in a high-energy unstable state. In contrast, when the proton occupies an adjacent outward-bending site, the rotational phases reinforce each other, and the lattice-distortion interaction can counteract the Coulomb repulsion, leading to an attractive and energetically stable proton pair. Therefore, protons tend to be trapped at adjacent inward-bending sites. In addition, the lowest-energy configuration of the proton-pair system corresponds to two protons located within the same Y-induced trapping plane, both forming hydrogen bonds with the same oxygen. In this structure, one proton occupies another type of trapping site, where it is cooperatively stabilized by both the other proton and the Y dopant. To investigate the impact of proton pairing on conduction, we computed long-range diffusion pathways under both single- and two-proton conditions across various local environments. The resulting range of rate-limiting barriers is 0.24–0.45~eV for two-proton conduction and 0.19–0.39~eV for single-proton conduction. The higher, experimentally comparable barriers in the two-proton case all correspond to rotations or transfers in which one proton resides at a site trapped by another proton. This indicates that the proton trapping effect induced by pairing hinders proton conduction. Our study elucidates the multi-proton diffusion mechanism and its impact on proton transport, providing a theoretical foundation for the experimental design of electrolyte materials with high proton conductivity.

\noindent
\underbar{\bf Acknowledgements:}
This work was supported by National Natural Science Foundation of China (Nos. 12404463 and 12474218), and Beijing Natural Science
Foundation (Nos. 1242022 and 1252022),  the Double First-Class Construction Fund for Teacher Development Projects (0515024GH0201201 and 0515024SH0201201), and the Key Research and Development Program of Shaanxi (2024GX-YBXM-456 and 2024GX-YBXM-565).

\noindent
\underbar{\bf Data availability:}
The data that support the findings of this article are openly available\cite{ma_2025_17506012}.

%%%%%%%%%%%%%%%%%%%%%%%%%%%%%%%%%%%%%%%%%%%%%%%%%%%%%%%%%%%%%%%%%%%%%%%%
%%%%%%%%%%%%%%%%%%%%%%%%%%%%%%%%%%%%%%%%%%%%%%%%%%%%%%%%%%%%%%%%%%%%%%%%
%Supplemental Material
%%%%%%%%%%%%%%%%%%%%%%%%%%%%%%%%%%%%%%%%%%%%%%%%%%%%%%%%%%%%%%%%%%%%%%%%
%%%%%%%%%%%%%%%%%%%%%%%%%%%%%%%%%%%%%%%%%%%%%%%%%%%%%%%%%%%%%%%%%%%%%%%%

\appendix
\section*{Appendix}

\setcounter{equation}{0}
\setcounter{figure}{0}
\renewcommand{\theequation}{A\arabic{equation}}
\renewcommand{\thefigure}{A\arabic{figure}}
\renewcommand{\thesubsection}{A\arabic{subsection}}

\subsection*{1. Calculation procedure and data for the proton-proton interaction energy}

\begin{figure}[htbp]
\centering
\includegraphics[scale=0.253]{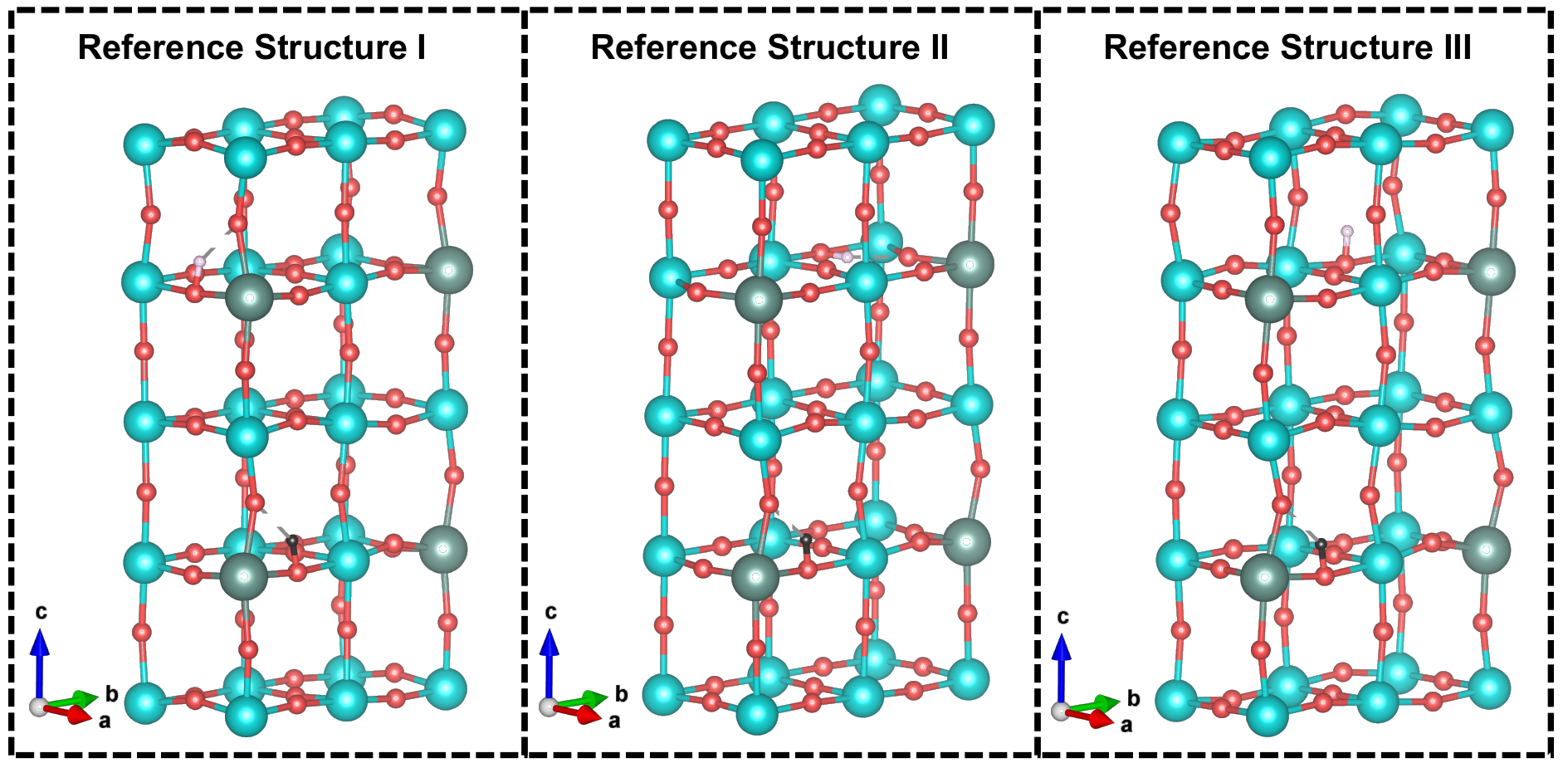}
\caption{Independent proton reference structures used to calculate the interaction energies of different types of two-proton configurations shown in Fig.~2(a) and (c).} 
\label{FigA1}
\end{figure}

\begin{table}[htbp]
\centering
\caption{Total energy ($E_{tot}$), independent proton reference energy ($E_{ref}$), interaction energy between protons ($E_{int}$), electrostatic repulsion ($E_{elec}$), and the elastic interaction ($E_{elas}$) for all two-proton configurations shown in Fig.~2(a) and (c).}
\label{TabA1}
\renewcommand{\arraystretch}{1.5}
\begin{tabularx}{\linewidth}{|c|X|X|X|X|X|}
\hline
\textbf{site} & \textbf{$E_{tot}$ (eV)} & \textbf{$E_{ref}$ (eV)} & \textbf{$E_{int}$ (eV)} & \textbf{$E_{elec}$ (eV)} & \textbf{$E_{elas}$ (eV)} \\
\hline
1    & -667.67    &         & -0.19    & 0.57    & -0.76    \\
3    & -667.48    &         & 0        & 0.36    & -0.36    \\
5    & -667.61    & -667.48 & -0.13    & 0.12    & -0.25    \\
7    & -667.47    &         & 0.0      & 1.3     & -1.3     \\
9    & -667.41    &         & 0.07     & 0.88    & -0.81    \\
11   & -667.64    &         & -0.16    & 0.33    & -0.49    \\
\hline 
4    & -667.57    &         & -0.11    & 0.14    & -0.28    \\
6    & -667.17    & -667.46 & 0.29     & 0.025   & 0.265    \\
10   & -667.07    &         & 0.39     & 0.49    & -0.1     \\
12   & -667.57    &         & -0.11    & 0.14    & -0.28    \\
\hline 
2    & -667.4     & -667.33 & -0.07    & 0.42    & -0.49    \\
8    & -667.42    &         & -0.09    & 0.89    & -0.98    \\
\hline
13    & -666.81   & -667.15 & 0.35    & 0.21    &  0.14    \\
\hline
14    & -667.16   & -667.22 & 0.06    & 0.1     & -0.04    \\
\hline
\end{tabularx} 
\end{table}

Figure~\ref{FigA1} shows the structures used as independent proton reference configurations, in which the two protons are placed in different environments and are sufficiently far apart to be considered non-interacting. Reference Structure I corresponds to configurations where both protons occupy nearest-neighbor sites of a Y dopant (corresponding to proton 2 at sites 4, 6, 10, and 12 in Fig.~\ref{Fig2}(a)(c)). Reference Structure II represents configurations where the proton 2 are located at next-nearest-neighbor sites, oriented toward the Y-containing plane (corresponding to sites 1, 3, 5, 7, 9, and 11). Reference Structure III corresponds to configurations where the proton 2 are also at next-nearest-neighbor sites but oriented away from the Y-containing plane (corresponding to sites 2 and 8). For each reference configuration, only a representative structure is shown here. In practice, we performed convergence tests on the total energy for each type of reference configuration to ensure that the two protons are sufficiently far apart to be regarded as non-interacting. For example, the converged energies for Reference Structures I, II, and III are -667.46~eV, -667.48~eV, and -667.33~eV, respectively (TABLE.~\ref{TabA1}), with the corresponding minimum proton-proton distances being 8.90~\AA, 8.33~\AA, and 9.08~\AA.

To verify that a proton can influence only another proton located in the same plane via lattice distortion, we additionally calculated configurations where proton 2 occupies sites 13 and 14, i.e., not in the same plane as proton 1. In this case, the interaction energies $E_{int}$ between the two protons for configurations 13 and 14 are 0.35 eV and 0.06 eV, respectively, indicating that proton pairs cannot be formed. This is mainly because the Coulomb repulsion $E_{elec}$ dominates (0.21 eV and 0.10 eV, respectively), while the lattice distortion interaction is relatively weak, and thus the Coulomb repulsion cannot be compensated to form a lattice-distortion-mediated proton pair.

\subsection*{2. Periodic boundary condition test for lattice distortions induced by proton}

\begin{figure}[htbp]
\centering
\includegraphics[scale=0.43]{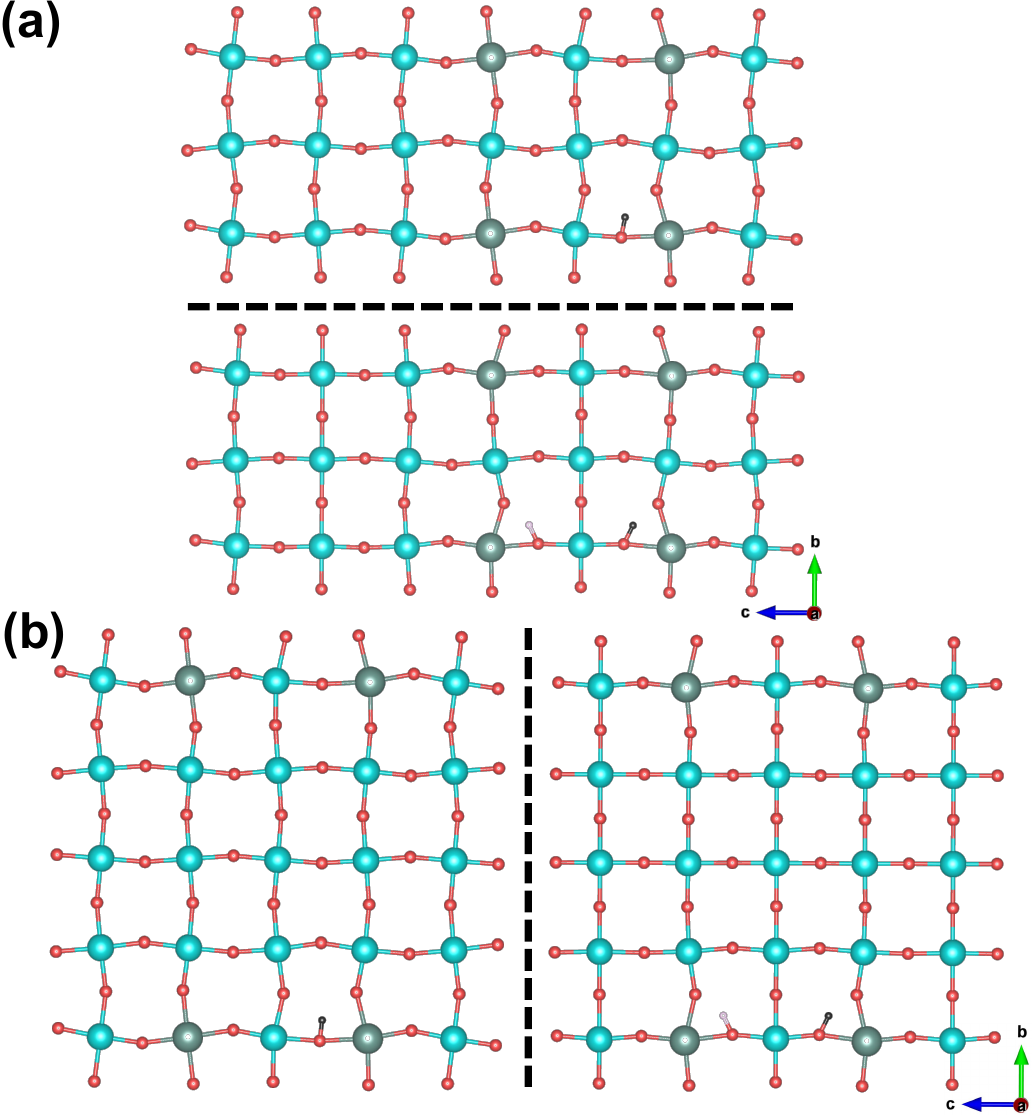}
\caption{Schematic illustrations showing that proton 2 at the inward-bending repulsive site disrupts the antiferrodistortive octahedral rotation in the (a) $2\times2\times6$ and (b) $4\times4\times4$ supercells.} 
\label{FigA2}
\end{figure}

To rule out artifacts from periodic boundary conditions in our $2\times2\times4$-based analysis, we also examined the lattice distortions induced by protons in larger $2\times2\times6$ and $4\times4\times4$ supercells. As shown in the upper panel of Fig.~\ref{FigA2}(a) and the left panel of Fig.~\ref{FigA2}(b), proton 1 induces antiferrodistortive octahedral rotation within its hosting plane. However, when proton 2 occupies the inward-bending repulsive site generated by proton 1, these antiphase rotations are suppressed, as illustrated in the lower panel of Fig.~\ref{FigA2}(a) and the right panel of Fig.~\ref{FigA2}(b).

\subsection*{3. The interactions between protons under non-uniform distribution of Y dopants}

\begin{figure}[htbp]
\centering
\includegraphics[scale=0.37]{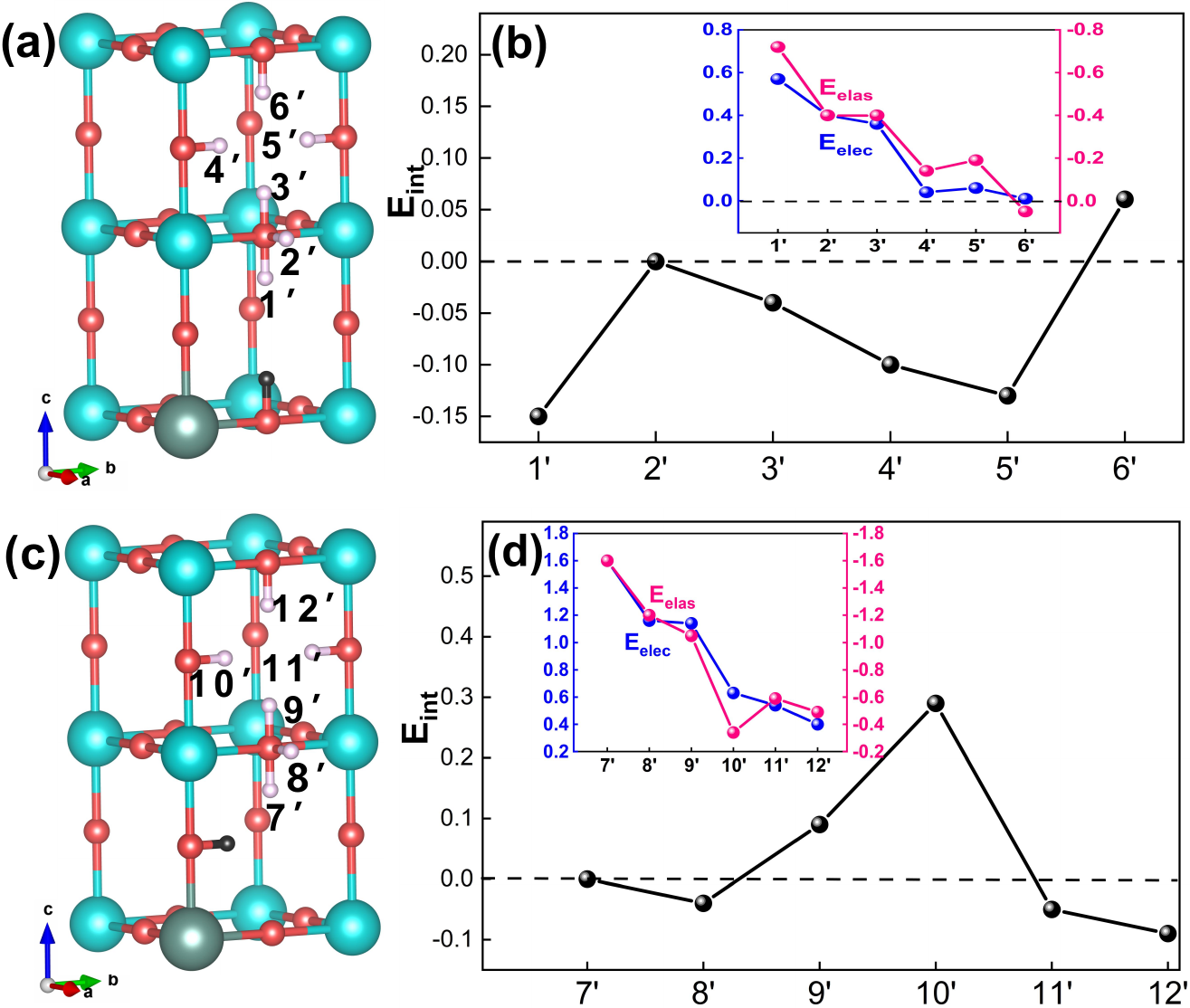}
\caption{(a,c) Schematic illustration of two-proton configurations with proton 1 fixed at the Y-adjacent oxygen site along the b-axis (a) and c-axis (c), in the absence of Y dopants in the neighboring unit cell of proton 1. The black sphere represents the fixed proton 1, and the white sphere represents proton 2 at different positions. (b,d) Proton-proton interaction energy $E_{int}$, and electrostatic repulsion energy $E_{elec}$ (the blue line of insert) and elastic interaction energy $E_{elas}$ (the pink line of insert) for the configurations in (a) and (c), respectively. } 
\label{FigA3}
\end{figure}

\begin{table}[htbp]
\centering
\caption{Total energy ($E_{tot}$), independent proton reference energy ($E_{ref}$), interaction energy between protons ($E_{int}$), electrostatic repulsion ($E_{elec}$), and the elastic interaction ($E_{elas}$) for all two-proton configurations shown in Fig.~\ref{FigA3}(a,c).}
\label{TabA2}
\renewcommand{\arraystretch}{1.5}
\begin{tabularx}{\linewidth}{|c|X|X|X|X|X|}
\hline
\textbf{site} & \textbf{$E_{tot}$ (eV)} & \textbf{$E_{ref}$ (eV)} & \textbf{$E_{int}$ (eV)} & \textbf{$E_{elec}$ (eV)} & \textbf{$E_{elas}$ (eV)} \\
\hline
1$'$    & -667.64    & -667.49    & -0.15    & 0.57    & -0.72    \\
7$'$    & -667.45    &            & 0.0      & 1.6     & -1.6     \\
\hline
2$'$    & -667.3     & -667.34    & 0.0      & 0.4     & -0.4     \\
8$'$    & -667.38    &            & -0.04    & 1.16    & -1.2     \\
\hline
3$'$    & -667.45    & -667.41    & -0.04    & 0.36    & -0.4     \\
9$'$    & -667.32    &            & 0.09     & 1.14    & -1.05    \\
\hline
4$'$    & -667.5     &            & -0.1     & 0.04    & -0.14    \\
5$'$    & -667.53    & -667.4     & -0.13    & 0.06    & -0.19    \\
10$'$   & -667.11    &            & 0.29     & 0.63    & -0.34    \\
11$'$   & -667.45    &            & -0.05    & 0.54    & -0.59    \\
\hline
6$'$    & -667.27    & -667.33    & 0.06     & 0.01    & 0.05     \\
12$'$   & -667.42    &            & -0.09    & 0.4     & -0.49    \\
\hline
\end{tabularx} 
\end{table}

To rule out the influence of uniformly distributed Y dopants on the net repulsive two-proton configurations, we also examined configurations under a non-uniform 12.5\% Y doping condition. As shown in Fig.~\ref{FigA3}, no Y dopants are present in the neighboring unit cell where proton 1 resides. In this case, proton 2 still exhibits a clear net repulsive interaction with proton 1 when located at the same positions (6$'$ and 10$'$). The similar trends observed in $E_{int}$ and $E_{elas}$ indicate that the elastic interaction remains the dominant factor governing whether the two protons exhibit net attraction or repulsion. The two-proton configurations with one proton at site 6$'$ or 10$'$ also exhibit significantly elongated $H-O_{nn}$ bond lengths, measured to be 2.81~\AA\ and 3.05~\AA, respectively, compared to 2.28~\AA\ and 2.25~\AA\ in the corresponding single-proton cases. These elongated bonds destabilize the proton, and imaginary phonon frequencies are observed in both cases. The increase in $H-O_{nn}$ bond length is again attributed to the excess proton at the inward-bending sites induced by proton 1, which disrupts the antiferrodistortive octahedral rotation originally induced by the single proton.

\subsection*{4. Statistics of the spatial correlation between two protons based on AIMD simulations}

\begin{figure}[htbp]
\centering
\includegraphics[scale=0.35]{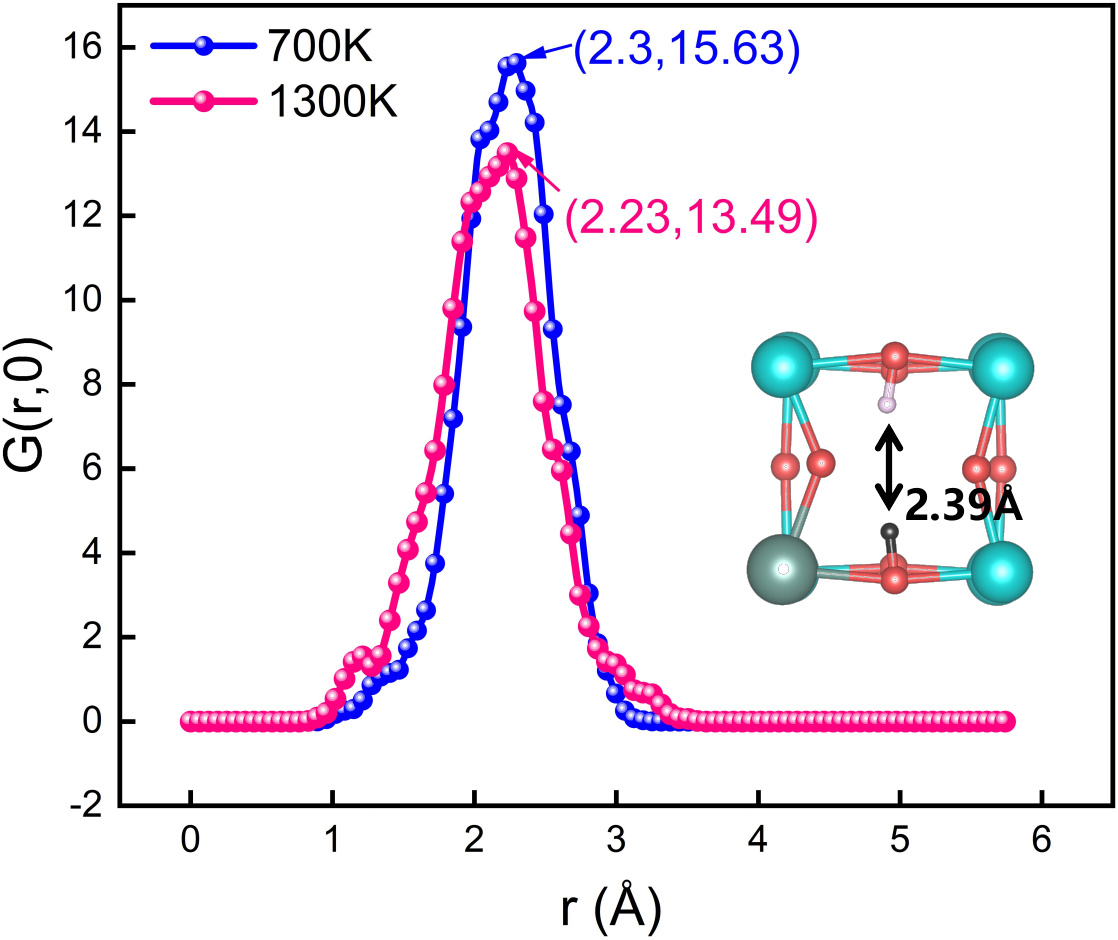}
\caption{Static radial distribution functions between two protons at 700~K and 1300~K. The insert is the schematic illustration of the lowest-energy two-proton configuration based on DFT.} 
\label{FigA4}
\end{figure}

To further examine whether two protons tend to be trapped at the sites corresponding to the lowest-energy proton pair configuration identified by our DFT calculations, we performed 10 ps ab initio molecular dynamics (AIMD) simulations for $3 \times 3 \times 3$ BaZrO$_3$ system with 3.7\% Y doping containing two protons at 700~K and 1300~K. The simulations were carried out for 10~ps with a time step of 0.333~fs under the NVT ensemble using a Nos\'e-Hoover thermostat\cite{PhysRevB.33.8822,10.1063/1.2130390}. The $\Gamma$-point was used for Brillouin zone sampling during the AIMD simulations. To investigate the space-time correlation between two protons, we analyzed their positions using the van Hove correlation function (Eq.~\eqref{Equ1}), specifically its distinct part\cite{PhysRev.95.249}.

\begin{equation}
    \label{Equ1}  
     G_{d}(r, \Delta t) \propto \sum_{a \neq b}^{N} \delta \bigl( r - | r_{a}(\Delta t) - r_{b}(0) | \bigr) 
\end{equation}

Here, $r_{a}(\Delta t)$ and $r_{b}(0)$ denote the positions of particle a at time t and particle b at the initial time, respectively. The $G_{d}(r, \Delta t)$ function describes the probability density of finding particle a at position $r_{a}(\Delta t)$ after time $\Delta t$, given that particle b was at position $r_{b}(0)$ at the initial time. This captures the space-time correlation between two particles. The statistical results show no stable pattern of sequential or simultaneous transfer between the two protons. Therefore, we focus on the spatial correlation at $\Delta = 0$, which corresponds to the static radial distribution function between the two protons. As shown in Fig.~\ref{FigA4}, the peak positions of $G_{d}(r, 0)$ at 700~K and 1300~K are 2.30~\AA\ and 2.23~\AA, respectively, which are close to the proton-proton distance in the lowest-energy two-protons configuration. This indicates that two protons localized near the Y dopant tend to adopt the lowest-energy configuration, in which one proton is simultaneously trapped by both the Y dopant and the other proton.

\subsection*{5. Two Y-doping configurations at 12.5\% concentration}

\begin{figure}[htbp]
\centering
\includegraphics[scale=0.37]{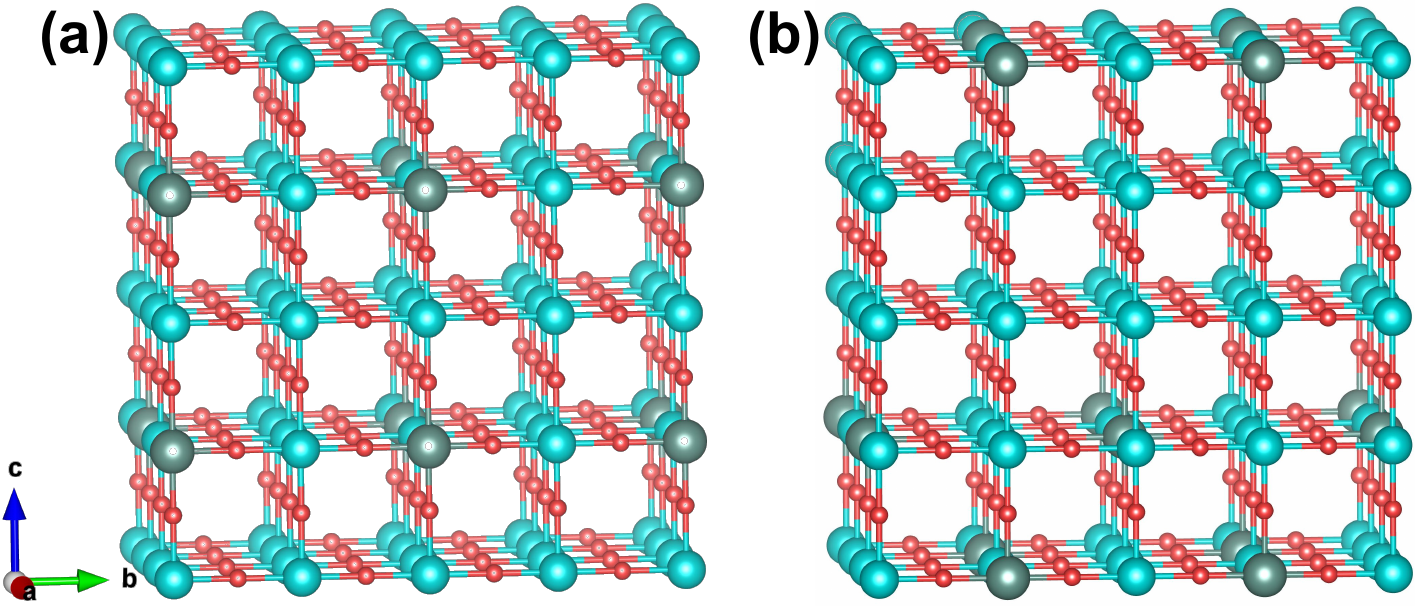}
\caption{Two cases of calculated long-range proton migration barriers in 12.5\% Y-doped BaZrO$_3$: (a) uniformly distributed Y; (b) non-uniformly distributed Y.} 
\label{FigA5}
\end{figure}

To better reflect realistic conditions, we calculated the long-range proton migration barriers under two different Y doping configurations. As shown in Fig.~\ref{FigA5}(a), Y dopants are uniformly distributed along the a, b, and c directions with one Zr separation between adjacent dopants. To create a non-uniform distribution, one layer of Y dopants was sequentially shifted by one lattice constant along the c, b, and a directions, resulting in the configuration shown in Fig.~\ref{FigA5}(b). The non-uniform distribution includes two types of Y arrangements: a low-density distribution, in which adjacent Y are located in next-nearest-neighbor unit cells and exhibit the furthest diagonal separation; and a high-density distribution, where adjacent Y occupy diagonal sites within the same unit cell.

\subsection*{6. The computational procedure of the minimum-energy two-proton pathway}

\begin{figure*}[htbp]
\centering
\includegraphics[scale=0.47]{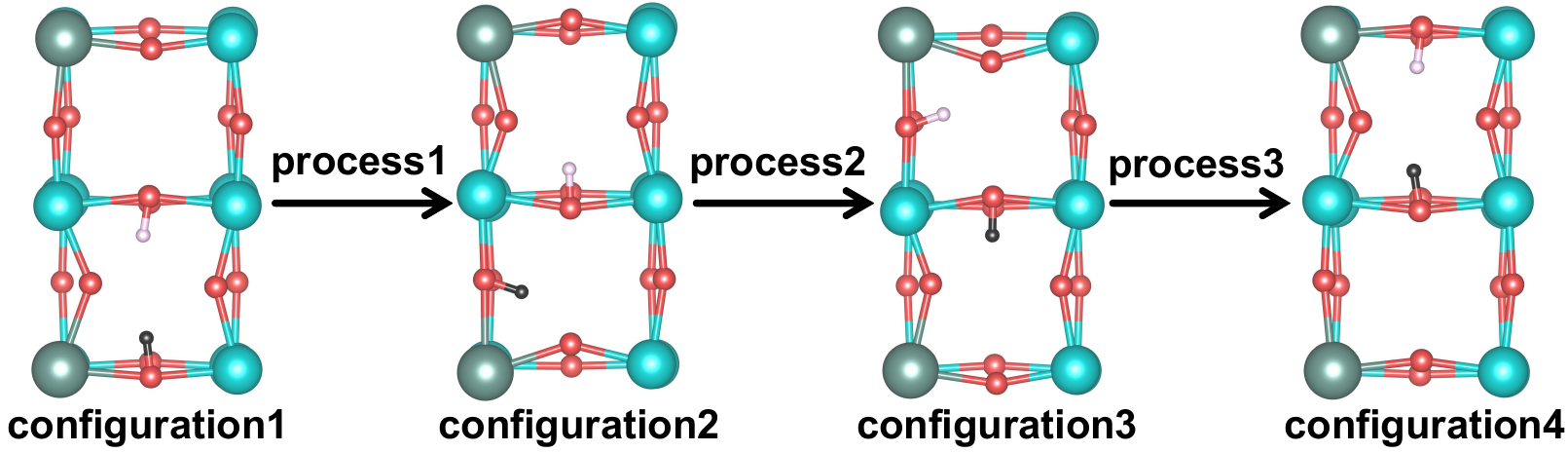}
\caption{ Schematic illustration of the method used to determine the minimum energy pathway for two-proton migration based on CI-NEB calculation.} 
\label{FigA6}
\end{figure*}

\begin{table}[htbp]
\centering
\caption{All possible sequential two-proton migration cases and their corresponding energy barriers for each intermediate process shown in Fig.~\ref{FigA6}.}
\label{TabA3}
\renewcommand{\arraystretch}{1.5}
\begin{tabularx}{\linewidth}{|c|X|X|}
\hline
\textbf{} & \textbf{sequential migration path} & \textbf{the energy barrier (eV)}   \\
\hline
             & $t1 \rightarrow r1 \rightarrow r2$    & $0.22 \rightarrow 0.34 \rightarrow 0.12$  \\
process 1    & $r1 \rightarrow t1 \rightarrow r2$    & $0.27 \rightarrow 0.23 \rightarrow 0.12$   \\
             & $r1 \rightarrow r2 \rightarrow t1$    & $0.27 \rightarrow 0.17 \rightarrow 0.14$    \\
\hline 
process 2    & $t2 \rightarrow t3$   & $0.34 \rightarrow 0$   \\
\hline 
process 3    & $t4 \rightarrow r3 \rightarrow r4$    & $0.07 \rightarrow 0.26 \rightarrow -0.39$    \\
             & $r3 \rightarrow t4 \rightarrow r4$    & $0.11 \rightarrow 0.2 \rightarrow 0.39$    \\
             & $r3 \rightarrow r4 \rightarrow t4$    & $0.11 \rightarrow 0.12 \rightarrow 0.02$   \\
\hline
\end{tabularx} 
\end{table}

Figure~\ref{FigA6} illustrates the computational procedure used to obtain the minimum-energy two-proton migration pathway shown in Fig.~\ref{Fig4}. Starting from the initial and final lowest-energy configurations (configuration 1 and configuration 4), nine intermediate images were inserted and a rough preliminary CI-NEB calculation was performed. This yielded two converged local minima-configuration 2 and configuration 3. Based on these intermediate and endpoint configurations, the entire migration pathway was divided into three segments: process 1, process 2, and process 3. Considering sequential migrations of the two protons, each process involves multiple possible migration sequences. We systematically evaluated all possible two-proton migration sequences for each process by performing detailed CI-NEB calculations for each case. The minimum-energy pathway was then constructed by selecting, for each process, the sequence with the lowest rate-limiting barrier. For example, as shown in Table~\ref{TabA3}, process 1 involves a transfer t1 of proton 1 and two rotations (r1 and r2) of proton 2. Considering that only t1 or r1 can initiate the sequence, we identified three possible cases: $t1 \rightarrow r1 \rightarrow r2$, $r1 \rightarrow t1 \rightarrow r2$ and $r1 \rightarrow r2 \rightarrow t1$. We computed the migration barriers in the given order for each case, noting that the barrier for the same step may differ depending on its position in the sequence. Among the three, $r1 \rightarrow r2 \rightarrow t1$ yielded the lowest rate-limiting barrier and was thus selected as the $R1 \rightarrow R2 \rightarrow T1$ path in Fig.~\ref{Fig4}. Process 2 involves one transfer each of proton 1 and proton 2 (t2 and t3, respectively). As two protons cannot occupy the same oxygen site simultaneously, only one valid sequence exists-$t2 \rightarrow t3$-which corresponds to the $T2 \rightarrow T3$ segment in Fig.~\ref{Fig4}. Process 3 includes two rotations (r3 and r4) of proton 1 and one transfer (t4) of proton 2. By considering only sequences starting with either r3 or t4, we obtained three cases: $t4 \rightarrow r3 \rightarrow r4$, $r3 \rightarrow t4 \rightarrow r4$ and $r3 \rightarrow r4 \rightarrow t4$. Among these, the lowest rate-limiting barrier was found for $r3 \rightarrow r4 \rightarrow t4$, which was thus selected as the $R3 \rightarrow R4 \rightarrow T4$ portion of the final minimum-energy path.

\bibliography{ref}

\end{document}